# Correction Four-Component Dirac-Coulomb Using Gaussian Basis-Set and Gaussian Model Distribution for Super Heavy Element (Z=115)

Bilal K. Jasim
Ayad A. Al-Ani
Saad N. Abood

*Department of Physics,
College of Science,
Al-Nahrain University,
Baghdad, IRAQ*

*In this paper, we consider the Dirac-Coulomb equation for many-particles, to describe the interaction between electrons in the system having many electrons. The four-component wave function will expanding into a finite basis-set, using Gaussian basis function technique, in order to describe the upper and lower two components of the 4-spinors, respectively. Gaussian basis-set type dyall.v2z has been adopted to describe the correlation and polarization of 4-component wave function. The small component Gaussian basis functions have been generated from large component Gaussian basis functions using kinetic balance relation. The considered techniques have been applied for super heavy element $^{115}$Uup, in which the nuclei has large charge and the inner spinors $s_{1/2}$ is strongly contracted. To solve the problem resulted from singularity at the origin for the $1s_{1/2}$ spinors. We adopting the Gaussian charge distribution model to describe the charge of nuclei. To calculate accurate properties of the atomic levels, we used Dirac-Hartree-Fock method, which have more flexibility through Gaussian basis-set to treat relativistic quantum calculation for system has many-particle.*

## 1. Introduction

In most quantum calculations based on expansion methods, the nuclei charge distributions are described by a point charge. The Dirac equation for a single electron in the field of point charge can be solved analytically [1]. The resulting relativistic for $1S_{1/2}$ Dirac-Coulomb wave function has singularities at the origin. This problem can be diminished by the use of a more realistic finite nuclear charge distribution, i.e. Gaussian distribution [2]. We describe the status of the problem of the electron structure of the super heavy atom with nuclear charge Z=115 [3]. By using a basis-set in calculation on heavy element consist for computational reasons of Gaussian functions and it is difficult to describe a function with a non-zero derivative at the origin with such a basis-set. The Dirac-Coulomb Hamiltonian form a natural starting point for relativistic atomic methods. This operator leads, however, to a wave function constructed from complex 4-spinors [4]. The four-component wave function will expanding into a finite basis-set, by using Gaussian basis functions to describe 4-spinors. The use of Gaussian type Dyall basis-set in relativistic Dirac-Hartree-Fock calculations is likely to prove more difficult than in the corresponding non-relativistic cases. The cusp a point nucleus is infinite in relativistic wave functions, thus more difficult to represent with a basis-set expansion [5]. Relativistic effect is most important in heavy atoms. It will be necessary to treat these species with Dirac-Hartree-Fock basis-set expansion calculations. The treatment of some relativistic effect requires an accurate description of the wave function in the inner core origin.

## 2. Theory

The start point for relativistic atomic methods is the Dirac Coulomb equation can be written in atomic units i.e. ($\hbar=m=e^2/4\pi\varepsilon_0=1$) is given by [6]:

$$\left(\sum_{i=1}^{N}[c\hat{\alpha}_i \cdot \hat{p}_i + (\hat{\beta}_i - 1)c^2 + \hat{V}(r)] + \sum_{i<j}^{N}\hat{V}(i,j)\right)\psi_{n\kappa m} = E\psi_{n\kappa m} \quad (1)$$

where c is the speed of light (c=137.03602) in atomic unit, $\alpha = (\alpha_x, \alpha_y, \alpha_z)$ and β are the 4×4 Dirac matrices are given by [7]:

$$\alpha_x = \begin{pmatrix} 0_2 & \sigma_x \\ \sigma_x & 0_2 \end{pmatrix}, \alpha_y = \begin{pmatrix} 0_2 & \sigma_y \\ \sigma_y & 0_2 \end{pmatrix}, \alpha_z = \begin{pmatrix} 0_2 & \sigma_z \\ \sigma_z & 0_2 \end{pmatrix}, \beta = \begin{pmatrix} I_2 & 0_2 \\ 0_2 & -I_2 \end{pmatrix} \quad (2)$$

$\sigma_x, \sigma_y$, and $\sigma_z$ are the Pauli matrices, are given by [8]

$$\sigma_x = \begin{pmatrix} 0 & 1 \\ 1 & 0 \end{pmatrix}, \sigma_y = \begin{pmatrix} 0 & -i \\ i & 0 \end{pmatrix}, \sigma_z = \begin{pmatrix} 1 & 0 \\ 0 & -1 \end{pmatrix} \quad (3)$$

where $I_2$ and $0_2$ are 2×2 unit and zero matrices, respectively

The term $(\beta_i - 1)$ appearing in Dirac-Coulomb equation (1) to align the energy scale with the non-relativistic energy scale by subtracted the electron rest energy, i.e., replacing $\beta$ by $(\beta_i - 1)$. $\hat{p} = -i\nabla$ is the momentum operator, $\hat{V}(r)$ is the Coulomb potential for one electron, in this work we adopted the Gaussian distribution model to describe the nucleus charge. The Gaussian nuclear charge distribution is given by [8]

$$\rho_N(r_i) = Z_N \left(\frac{\eta_N}{\pi}\right)^{3/2} exp\left(\eta_N r^2_{iN}\right) \quad (4)$$

where Z is the nuclear charge and the exponent of the normalization Gaussian type function represent the nuclear distribution, determined by the root-mean-square radius of the nuclear charge distribution via the relation given by

$$\eta = \frac{3}{2\langle r^2\rangle} \quad (5)$$

where $\eta$ is the exponential parameter choosing to give a root-mean-square value. The potential $\hat{V}(r)$ in Eq. (1) for this charge density distribution (Gaussian model) is given by [9]

$$\hat{V}(r_i) = \sum_{i=1}^{N} \int \frac{\rho_N(r_I)}{|r_i-r_I|} dr_I \quad (6)$$

where $\rho_N(r_I)$ represents the nuclear charge distribution. $\hat{V}(i,j)$ represent the interaction energy of electron i and j, $\hat{V}(i,j) = \frac{1}{r_{ij}}$, $r_{ij} = |r_i - r_j|$.

$\psi_{n\kappa m}$ is the 4-spinor structure is given by in case of a spherical potential [10]

$$\psi_{n\kappa m} = \begin{pmatrix}\psi^L\\\psi^S\end{pmatrix} = \begin{pmatrix} u_{n\kappa}(r) & \chi_{jm\kappa}(\theta,\varphi)\\ -v_{n\kappa}(r) & \chi_{jm-\kappa}(\theta,\varphi)\end{pmatrix} \quad (7)$$

where $\psi^L$ and $\psi^S$ represents are the large and small components, respectively, $u_{n\kappa}(r)$ and $v_{n\kappa}(r)$ are the radial functions, $\chi_{jm\kappa}$ and $\chi_{jm-\kappa}$ are two-component spinors

The radial functions $u_{n\kappa}(r)$ and $v_{n\kappa}(r)$ may be expanded in a basis-set using Gaussian basis-set given by [11].

$$u_{n\kappa}(r) = \sum_i^N f^L_{\kappa p}(r) \xi_{n\kappa p} \quad (8)$$
$$v_{n\kappa}(r) = \sum_i^N f^S_{\kappa q}(r) \eta_{n\kappa q} \quad (9)$$

where $\xi_{n\kappa p}$ and $\eta_{n\kappa q}$ are linear variation parameters, $f^L_{\kappa p}(r)$ and $f^S_{\kappa q}(r)$ are the Gaussian basis-set for large and small components, respectively, given by [12]

$$f^L(r) = N_L r exp(-\zeta_L) r^2 \quad (10)$$
$$f^S(r) = N_S r exp(-\zeta_S) r^2 \quad (11)$$

The factors $\zeta_L$ and $\zeta_S$ in the exponents are the only adjustable parameters of these basis functions and these parameters $\zeta_L$ and $\zeta_S$ usually called the exponents of the basis function. $N_L$ and $N_S$ are normalization factors. Substituting Eq. (10) and Eq. (11) into Eq. (8) and (9), respectively, the radial functions become

$$u(r) = \sum_{i=1}^{N} r \, exp(-\zeta_L r^2) \, \xi_i \quad (12)$$
$$v(r) = \sum_{i=1}^{N} r \, exp(-\zeta_S r^2) \, \eta_i \quad (13)$$

where $\xi_i$ and $\eta_i$ are the linear variation parameters, N is the expansion length. The exponential factors $\zeta_L$ and $\zeta_S$ are expressed as a geometric progression.

The strategy of the Dirac-Hartree-Fock (DHF) approach for calculating the electronic structure of atoms is setup an expansion for expectation value of the Dirac-Coulomb Hamiltonian for many-electron atoms. The Dirac-Coulomb energy of a four-component wave function can be written after mathematical processors as shown in Appendix.

The terms inside the first summation in Eq. (14) represent the total energy for one electron has only one occupied shell, $F_l(a,a)$, $G_l(a,b)$ are the radial integrals and $\Gamma^l_{j_a j_b}$ represent the coefficient of Clebsh-Gordan, and the terms in the second summation represent the total exchange energy, the factor $\frac{n_a-1}{2j_a}$ after multiplying exchange energy give total exchange energy between one electron and the electrons in different shell. The terms inside last summation represent the total Coulomb energy for an atom, the factor $\frac{1}{2}n_a(n_a - 1)$ represent the number of pairs electrons in a-shell.

## 3. Calculation and Results

The Dirac-Hartree-Fock radial functions of the shells occupied in the ground state, determine for the natural atom, Z=115. The large components $u(r)$ of these shells and the small components $v(r)$ are depicted in the graphs for the atom $^{115}$Uup. The large and small radial function described by Gaussian-Dyall basis set function type (dyall.v2z). Figure (1) represents the large components for all orbitals of $^{115}$Uup atom and figure (2) shows the magnification large radial functions in Fig. (1).

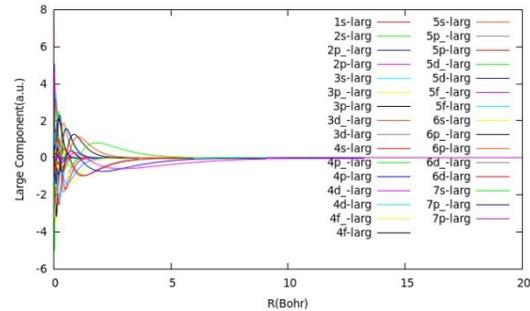

**Fig. (1) The large radial functions in atomic unit against R(Bohr) for all orbitals to Uup-atom using Gaussian-dyall.v2z basis-set**

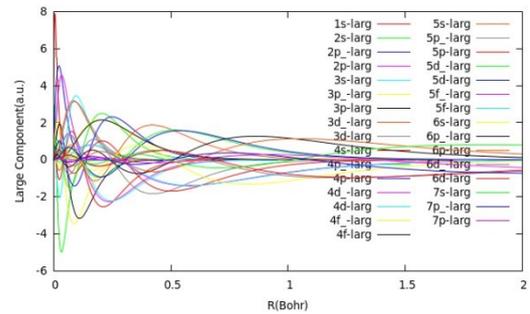

**Fig. (2) Magnification the large radial functions against R(Bohr) for all orbitals to Uup-atom**

Figure (3) shows the small component radial functions for all orbitals of $^{115}$Uup atom. It is clear that the small components radial functions are more compact and short range than the large component functions in Fig. (1). Figure (4) represents the magnification of the small component radial functions**.**

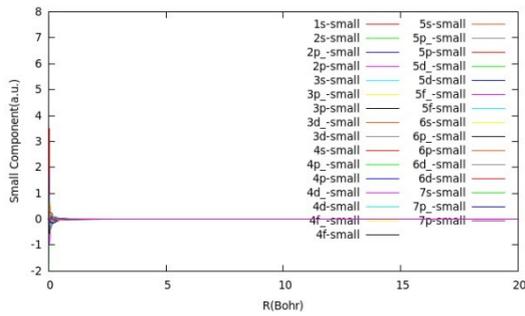

**Fig. (3) The large radial functions against R(Bohr) for all orbitals to Uup-atom using Gaussian-dyall.v2z basis-set**

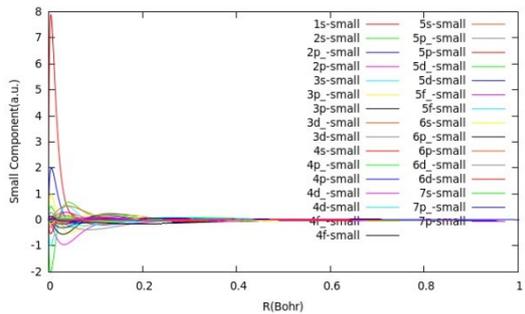

**Fig. (4) Magnification the small radial functions against R(Bohr) for all orbitals to Uup-atom**

Figure (5) explains the radial wave functions for the large components and small components for the $1S_{1/2}$ of the super heavy atom $^{115}$Uup. In calculations on $S_{1/2}$ state, the large component $u(r)$ is expanded of the familiar 1S Gaussian type function, is given by:

$$u(r) = \sum_{i}^{N} r\, exp(-\zeta_i r^2)\xi_i$$

where $N$ is the number of 1S primitive Gaussian type function and the small component $v(r)$ be expanded to

$$v(r) = \sum_{i}^{N} r^2\, exp(-\zeta_i r^2)\eta_i$$

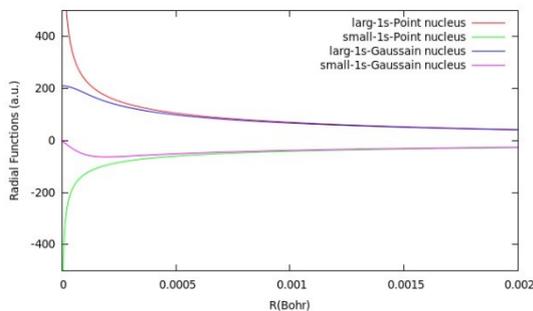

**Fig. (5) Comparison the radial wave functions $u(r)$ and $v(r)$ components versus R(Bohr) for point model and Gaussian model of super heavy element (Z=115) for closer orbital ($1S_{1/2}$)**

The both radial amplitudes large and small for $1S_{1/2}$ in Fig. (5) adopted point nucleus model and Gaussian nucleus model to describe the charge distribution system. At the center of the nucleus, the behavior of the point nuclear solution has singularity. In order to solve this problem we adopt, a Gaussian charge distribution model and Gaussian basis-set to gives a smooth and continuous wave function at the origin.

The effect of nuclear charge distribution on the spinor energy is notable when switching from the singular potential of point nucleus to Gaussian nucleus potential. The Gaussian nucleus potential not different very much, most important is the effect on relative energies. The spinor energies in Dirac-Hartree-Fock level, explain in table (1) for super heavy element (Z=115) in Hartree atomic unit. The results show between two different nuclear charge distribution models. Figure (6) shows the radial functions for $1S_{1/2}$ of the $^{115}$Uup-element, we note the magnitude of small component near to similar the large component but different sign.

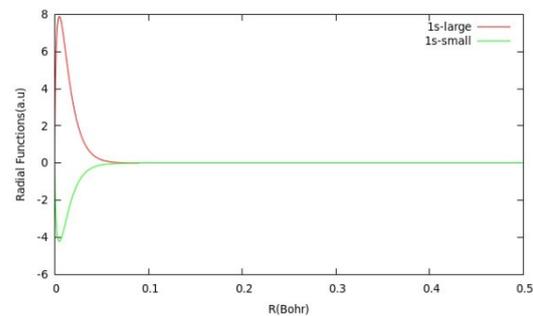

**Fig. (6) The radial wave functions $u(r)$ and $v(r)$ components versus R(Bohr) for Gaussian model of super heavy element (Z=115) for closer orbital ($1S_{1/2}$)**

### 4. Conclusion

For heavy atoms, the high nuclear charge creates a more pronounced cusp in case of non-relativistic when used point charge model at the origin for the $1S_{1/2}$. But the resulting in relativistic calculation appear singularity at the origin for the $1S_{1/2}$ and the solution is hard to describe with the conventional Slater function and even harder to describe with Gaussian functions. This means that we need much larger basis-set to solve this problem. Therefor a number of Gaussian type functions of high exponent must be included in a basis-set in order to mimic the region of the wave function near the origin. The super heavy element, such as: Z>115, the large component and small component for $1S_{1/2}$ orbital having similar absolute magnitude value but with negative sign at any distance r. For relativistic calculations, the point charge model not recommendable, especially, at or closer to the nuclei. This is because singularity appearance. Therefore, we adopted the Gaussian charge model combined with Gaussian basis function.

$$E_{DC} = \sum_a n_a \left( c\hbar \left( \int v_a(r) \left( \frac{\partial u_a(r)}{\partial r} + \frac{\kappa_a}{r} u_a(r) \right) dr - \int u_a(r) \left( \frac{\partial v_a(r)}{\partial r} - \frac{\kappa_a}{r} u_a(r) \right) dr \right) - \frac{Ze^2}{4\pi\varepsilon_0} \int \frac{1}{r} \left( u_a^2(r) + v_a^2(r) \right) dr + mc^2 \int \left( u_a^2(r) + v_a^2(r) \right) dr \right) - \sum_a \left( \frac{1}{2} n_a \frac{n_a-1}{2j_a} \sum_{l=0}^{\infty} \frac{1}{2}(2j_a+1) \Gamma_{j_a j_b}^l F_l(a,a) - \sum_{b \neq a} n_a \sum_{l=0}^{\infty} \frac{1}{4}(2j_b+1) \Gamma_{j_a j_b}^l G_l(a,b) \right) + \sum_a \left( \frac{1}{2} n_a (n_a - 1) F_0(a,a) + \frac{1}{2} \sum_{b \neq a} n_a n_b F_0(a,b) \right) \quad (14)$$

where $u(r)$ and $v(r)$ represent the radial components of the wave function

**Table (1) The relativistic spinor energy using different nuclear charge models for super heavy element (Z=115). Uunpentum atom Dyall basis set DZ have [26s 23p 17d 11f]**

| Level | Energy (a.u.) for Point model | Energy (a.u.) for Gaussian model | $E_G - E_p$ (a.u.) |
|---|---|---|---|
| 1s | -7692.74328 | -7608.83187 | 83.91140925 |
| 2s | -1594.729544 | -1575.873224 | 18.85632041 |
| 2p- | -1540.288406 | -1536.827733 | 3.460672528 |
| 2p | -1071.193542 | -1071.725444 | -0.531901317 |
| 3s | -433.9396493 | -429.2857713 | 4.653877957 |
| 3p- | -410.8960726 | -409.9316779 | 0.964394755 |
| 3p | -296.0324677 | -296.2084652 | -0.175997487 |
| 3d- | -265.662369 | -265.8271544 | -0.164785413 |
| 3d | -247.0539098 | -247.200756 | -0.146846198 |
| 4s | -127.9634222 | -126.608014 | 1.355408251 |
| 4p- | -117.2484671 | -116.9727943 | 0.275672765 |
| 4p | -83.93976753 | -84.00409862 | -0.064331095 |
| 4d- | -68.92886631 | -68.9863654 | -0.057499087 |
| 4d | -63.7507999 | -63.80278972 | -0.051989816 |
| 4f- | -44.11737616 | -44.16188326 | -0.044507092 |
| 4f | -42.59909669 | -42.6418525 | -0.042755805 |
| 5s | -35.1452495 | -34.7413907 | 0.403858805 |
| 5p- | -30.36977682 | -30.29379932 | 0.075977506 |
| 5p | -20.72638478 | -20.75055874 | -0.024173961 |
| 5d- | -14.16365862 | -14.18360113 | -0.019942513 |
| 5d | -12.8109854 | -12.82905899 | -0.018073597 |
| 5f- | -4.851061397 | -4.864141862 | -0.013080465 |
| 5f | -4.549872603 | -4.562437818 | -0.012565215 |
| 6s | -7.365786673 | -7.259823178 | 0.105963495 |
| 6p- | -5.542250521 | -5.525680465 | 0.016570056 |
| 6p | -3.298719414 | -3.307294525 | -0.008575111 |
| 6d- | -1.146712647 | -1.152217381 | -0.005504734 |
| 6d | -0.959372141 | -0.964003315 | -0.004631174 |
| 7s | -0.877225106 | -0.858372843 | 0.018852263 |
| 7p- | -0.447421946 | -0.445889515 | 0.00153243 |
| 7p | -0.213992145 | -0.215526459 | -0.001534314 |

Table (2) Comparison of the radial expectation value <R>, <1/R> and <R$^2$> between Gaussian model and point nucleus model of super heavy element (Z=115) using Dyall basis-set with Dirac-Hartree-Fock method

| | Point model | Gaussian model | Point model | Gaussian model | Point model | Gaussian model |
|---|---|---|---|---|---|---|
| level | <R> | <R> | <1/R> | <1/R> | <R$^2$> | <R$^2$> |
| 1s | 0.0091796 | 0.009315519 | 209.31907 | 199.58879 | 0.000124763 | 0.000127831 |
| 2s | 0.037747737 | 0.038157168 | 56.482765 | 54.011862 | 0.001768649 | 0.001804127 |
| 2p- | 0.029233786 | 0.029334117 | 55.779538 | 55.077457 | 0.00113792 | 0.001144451 |
| 2p | 0.043274943 | 0.043264269 | 29.666562 | 29.674043 | 0.00229133 | 0.002290204 |
| 3s | 0.10087824 | 0.10164956 | 19.352219 | 18.724111 | 0.01184166 | 0.012017581 |
| 3p- | 0.093496607 | 0.09368931 | 19.070599 | 18.871243 | 0.010392578 | 0.010432377 |
| 3p | 0.11847085 | 0.11843944 | 12.149627 | 12.152927 | 0.016395209 | 0.016386507 |
| 3d- | 0.10094203 | 0.10091404 | 12.023768 | 12.027196 | 0.011986442 | 0.011979822 |
| 3d | 0.1075556 | 0.10752876 | 11.085826 | 11.088579 | 0.013480584 | 0.013473829 |
| 4s | 0.21699424 | 0.21834565 | 8.0931133 | 7.8996009 | 0.053417037 | 0.05407387 |
| 4p- | 0.21325374 | 0.21358501 | 7.9140604 | 7.8529661 | 0.052060599 | 0.052216883 |
| 4p | 0.2576934 | 0.25761618 | 5.6187717 | 5.6204182 | 0.075588958 | 0.07554326 |
| 4d- | 0.2489391 | 0.24886099 | 5.4424725 | 5.4442275 | 0.071455093 | 0.071410127 |
| 4d | 0.26052017 | 0.26044253 | 5.1053595 | 5.1068906 | 0.078027519 | 0.077980942 |
| 4f- | 0.23962001 | 0.23954063 | 4.8858591 | 4.8874344 | 0.066131272 | 0.066086842 |
| 4f | 0.24497671 | 0.24489774 | 4.7614332 | 4.7629168 | 0.068968106 | 0.068922924 |
| 5s | 0.43002503 | 0.43251093 | 3.681542 | 3.6159538 | 0.20703427 | 0.20940676 |
| 5p- | 0.43789333 | 0.43850097 | 3.5402234 | 3.5206097 | 0.21563379 | 0.21621871 |
| 5p | 0.52270075 | 0.52249432 | 2.676657 | 2.6776364 | 0.3065967 | 0.30635 |
| 5d- | 0.54992911 | 0.54968884 | 2.47473 | 2.4758258 | 0.34245775 | 0.34215549 |
| 5d | 0.57443725 | 0.57419332 | 2.3377796 | 2.3388013 | 0.37328571 | 0.37296746 |
| 5f- | 0.65073705 | 0.65035196 | 1.9873625 | 1.9885229 | 0.4894226 | 0.48882957 |
| 5f | 0.66557784 | 0.66518347 | 1.9363584 | 1.9374819 | 0.51197351 | 0.51134987 |
| 6s | 0.86744182 | 0.87326472 | 1.6597404 | 1.6356449 | 0.83941396 | 0.85059181 |
| 6p- | 0.92297806 | 0.92450201 | 1.5364796 | 1.5299405 | 0.95426788 | 0.95734182 |
| 6p | 1.145176 | 1.1442491 | 1.1539922 | 1.1548279 | 1.4694408 | 1.4669412 |
| 6d- | 1.4147888 | 1.4129862 | 0.92232123 | 0.92348826 | 2.2863833 | 2.2802817 |
| 6d | 1.5111582 | 1.509236 | 0.85532781 | 0.85643528 | 2.6131314 | 2.6061828 |
| 7s | 2.0712314 | 2.0919357 | 0.63916057 | 0.63025309 | 4.8446051 | 4.9425795 |
| 7p- | 2.4979793 | 2.5041922 | 0.52153114 | 0.51958695 | 7.1249429 | 7.1600044 |
| 7p | 3.6225641 | 3.6095804 | 0.34899698 | 0.35015506 | 15.191185 | 15.077022 |